\documentclass[%
 reprint,
superscriptaddress,
 amsmath,amssymb,
 aps,
 prl,
]{revtex4-2}

\usepackage{graphicx}
\usepackage{dcolumn}
\usepackage{bm}
\usepackage{gensymb}
\usepackage{color}
\usepackage{ulem}

\usepackage{xcolor,soul}

\usepackage{hyperref}

\begin{document}

\preprint{APS/123-QED}

\title{Fractional Chern Insulators and Competing States in a Twisted MoTe$_2$ Lattice Model}
\author{Yuchi He}
	\email{yuchi.he@physics.ox.ac.uk (he/him/his)}
	\affiliation{Rudolf Peierls Centre for Theoretical Physics, University of Oxford, Oxford OX1 3PU, United Kingdom}
    	\affiliation{Department of Physics, Ghent University, Krijgslaan 281, 9000 Gent, Belgium}

 \author{S.H. Simon}
	\email{steven.simon@physics.ox.ac.uk}
	\affiliation{Rudolf Peierls Centre for Theoretical Physics, University of Oxford, Oxford OX1 3PU, United Kingdom}
	
	\author{S.A. Parameswaran}
	\email{sid.parameswaran@physics.ox.ac.uk}
	\affiliation{Rudolf Peierls Centre for Theoretical Physics, University of Oxford, Oxford OX1 3PU, United Kingdom}
		\date{\today}

\date{\today}

\begin{abstract}
We construct an interacting lattice model for  twisted  $\mathrm{MoTe}_{2}$ bilayers at a twist angle of approximately 3.7\degree. We use the infinite density matrix renormalization group (iDMRG) in a cylinder geometry to identify a variety of competing integer and fractional Chern insulators and charge density wave (CDW) states that emerge upon the spontaneous breaking of time reversal symmetry by valley polarization. We use finite-size analysis to establish the robustness of Chern insulating states even in geometries that admit  competing CDWs, and explore the phase transitions between these states driven by increasing sublattice potential or interaction strength. Our work highlights the crucial role played by direct spin exchange in stabilizing the parent valley-polarized Chern ferromagnet band, and by the mixing with higher bands in destabilizing CIs/FCIs in favor of CDW orders.
\end{abstract}

\maketitle

The observation of fractionally quantized anomalous Hall  plateaus in van der Waals materials~\cite{cai_signatures_2023,Zeng2023,Park2023,PhysRevX.13.031037,Lu2024} 
has initiated a new chapter in the study of the topological phases of matter. The two-dimensional electron gases (2DEGs) in these heterostructures provide a long-sought experimental platform for studying integer~\cite{PhysRevLett.61.2015,QAH} andfractional~\cite{sheng2011fractional,PhysRevLett.106.236804,PhysRevX.1.021014,PhysRevLett.106.236803,PhysRevLett.106.236802,Xiao2011} Chern insulators (CIs/FCIs), zero-field lattice analogs of the celebrated integer and fractional quantum Hall phases of 2DEGs in high magnetic fields.  FCIs have to date been reported in three distinct settings: magic-angle twisted bilayer graphene in a weak magnetic field; rhombohedrally-stacked crystallographic graphene multilayers aligned to a hexagonal boron nitride substrate; and twisted homobilayers of the transition metal dichalcogenide MoTe$_2$, our focus here.

The low-energy band structure of twisted MoTe$_2$ bilayers can be captured within a continuum model approximation that focuses on the moir\'e  reconstruction of the electronic states near $K$-points of the microscopic Brillouin zone. Spin-orbit coupling locks spin to valley, so that each low-energy band exhibits a twofold degeneracy when the approximate $U(1)_v$ valley conservation is promoted to an exact symmetry. For a range of twist angles, the kinetic energy is suppressed, enhancing the role of interactions in creating correlated states. Furthermore, the bands below the Fermi energy at charge neutrality are characterized by  non-zero Chern numbers $\pm C$, equal and opposite in the two valleys as required by time-reversal symmetry. These facts have motivated proposals that FCIs emerge via a two-step process, wherein interactions trigger the spontaneous breaking of time-reversal symmetry at integer filling $\nu=-1$  by polarizing electrons into a single valley (and hence spin) to form a CI, and subsequently stabilize FCIs when holes are doped into this topological band.

\begin{figure}[t!]
\begin{center}
\includegraphics[width=0.9\columnwidth]{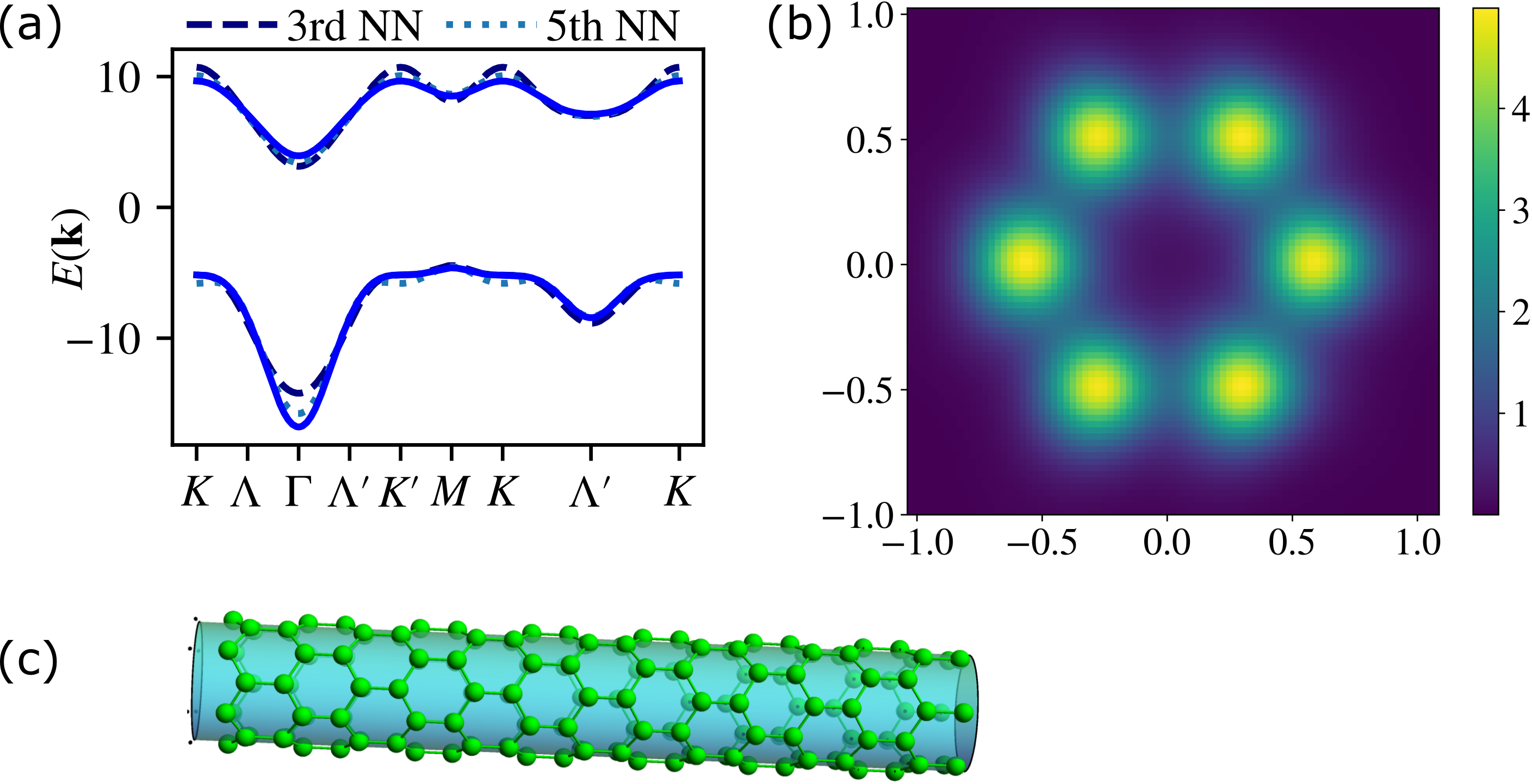}
\end{center}
\caption{
(a) Band dispersion: Top two valence bands of the continuum model (solid lines) and bands of Haldane model truncated to third and fifth nearest-neighbor hoppings. The unit of energy is meV. (b) Density of Wannier orbitals centered on a hexagon of the honeycomb lattice, constructed from the top two valence band (a) of the  continuum model for  DFT band structure~\cite{PhysRevLett.132.036501}.  
{The Wannier orbitals extend to both layers but mostly ($p\sim 0.93$) live on one layer for a given sublattice.   Only this larger component is plotted.}  The lattice constant is approximately $51.8\mathrm{\AA}$. The colorbar denotes the number density per unit cell.   (c) The zigzag nanotube YC6 geometry used in DMRG calculations. 
}\label{fig:Haldanesetup}
\end{figure}

While this scenario is superficially consistent with the observed phenomenology, there are several aspects of experiments and the band structure that hint at a more complicated picture. First, the precise sequence of the Chern numbers of successive low-energy bands is twist-angle dependent at small twist angles~\cite{park2024ferromagnetism,xu2025}, which is particularly important since the interaction scale is such that there is likely to be considerable mixing between these bands.  While there is consensus~\cite{PhysRevLett.132.036501,PhysRevB.109.205121,park2024ferromagnetism} that the two lowest-lying bands have $C=\pm1$ for the larger twist angles near $3.7$\degree~that we consider, even here there remain various puzzling features of the experiments. For instance, spin-valley-polarized FCIs are found at $\nu=-2/3,-3/5$, believed to emerge from a polarized CI at $\nu=-1$. However, such spin-valley ferromagnetism (as measured via optical absorption) is weaker at $\nu=-1/3$ than at $\nu=-2/3$, suggesting that the state is only partially polarized. Furthermore,  FCIs are observed in a relatively wide range of twist angles ($\sim 1 \degree$), hinting that the that fine-tuned flatness of electron dispersion is unnecessary. Indeed, a previous density functional theory (DFT) calculation~\cite{PhysRevLett.132.036501} for $3.89$\degree~ shows that the gap and the bandwidth of the first hole band are of comparable scale (Fig.~\ref{fig:Haldanesetup}(a)), so that interactions strong enough to stabilize FCIs likely also induce appreciable interband mixing, and is also in tension with the expectation that Chern ferromagnetism requires a high degree of band flatness. Experiments {at $\nu=-1/3$}~\cite{ji2024localprobebulkedge,xu2025signatures} also reveal the presence of a trivial correlated insulator, likely a charge-density wave (CDW). 
Understanding these features, as well as the stability of FCIs against CDWs in the broader parameter space of dielectric constant and gate-screening length (both of which tune the interactions) are  important aspects of the phenomenology that have yet to be fully explored.

Here, we address this challenge in the setting of an interacting lattice model for twisted MoTe$_2$ near a $3.7$\degree~twist angle. The central idea~\cite{PhysRevLett.122.086402,yu2020giant,Devakul2021,PhysRevX.13.041026} is to embed the low energy $C=\pm1$ bands in a Kane-Mele tight-binding model on the honeycomb lattice~\cite{PhysRevLett.95.146802}, whose parameters are fit to DFT data~\cite{PhysRevLett.132.036501}. Compactifying this lattice model into a cylinder geometry enables the application of large-scale infinite density-matrix renormalization group (iDMRG) simulations~\cite{White,McCulloch2008,10.21468/SciPostPhysLectNotes.5,iDMRGpaper} to study the emergence of CIs/FCIs and their stability against depolarization, band mixing, and competing CDWs.  This is in contrast to many previous works~\cite{li2021spontaneous,PhysRevB.107.L201109,PhysRevB.109.045147,PhysRevB.109.L121107,PhysRevResearch.5.L032022,abouelkomsan2022multiferroicity,PhysRevLett.132.036501,PhysRevB.108.085117,PhysRevLett.131.136501,PhysRevLett.131.136502,pnas.2316749121}, that variously either study a single band per valley or assume full spin polarization when studying FCIs, or study relatively small system sizes that may not accommodate competing CDWs due to commensuration effects.

In contrast to exact diagonalization --- utilized in many previous studies --- DMRG affords the study of geometries compatible with longer-wavelength candidate charge density waves, whose charge order can be directly measured in the quasi-one-dimensional thermodynamic limit. Furthermore, all spin sectors can be scanned rather than restricting to few spin flips atop a fullly-polarized phase, and correlation lengths can be estimated to check finite-size effects and assess the continuity of phase transitions. We also find that the short-range component of the Coulomb interaction, which manifests as a direct spin-exchange contribution within our lattice model and depends sensitively on details of the Wannier orbitals in the hole bands, appears to be crucial to quantitatively capture ferromagnetism. Our calculations suggest that dual-gate screening with a small gate distance makes FCIs less stable, and also predict that the CDW-FCI transitions could  be weakly first-order.

\noindent\textit{Model.---}The starting point of our study is a truncation to the Hilbert space of the low-energy bands.  We then construct a Hamiltonian in real space. This enables the study of band mixing and provides a real-space picture. A set of  DFT data~\cite{PhysRevLett.132.036501,zhang2024polarization} for the low energy bands of twisted bilayer MoTe$_2$ indicate angle-dependent Chern numbers consistent with experiments~\cite{cai_signatures_2023,Zeng2023,Park2023,PhysRevX.13.031037,Kang2024,park2024ferromagnetism,WSe2s,xia2024unconventional,knüppel2024correlated}.  Notably, near a twist angle of 3.89°, the top two valence bands for a given spin exhibit opposite Chern numbers ($\pm 1$)~\cite{PhysRevLett.132.036501}. This observation enables the representation of the two-band Hilbert space using Wannier orbitals which are exponentially localized on a honeycomb lattice, with moir\'e lattice constent $a \approx 51.8~{\rm \AA}$.
{We utilize the Wannierization scheme of Ref.~\cite{Devakul2021}.   The overlap of Wannier orbitals gives a hopping Hamiltonian 
\begin{equation}
H_0 =  \sum_{  i,j ,\sigma} t_{\bm{r}_i-\bm{r}_j,\sigma} c^{\dagger}_{i\sigma}c_{j\sigma}+\sum_{i}\mu_in_i, \ \label{eq:H0}
\end{equation}
where  $c^{\dagger}_{i\sigma}$ ($c_{i\sigma}$) annihilates (creates) one electron in the Wannier orbital on site $i$ of the honeycomb, and we keep hopping terms up to the third nearest neighbor (see Ref.~\cite{suppm}).    The hopping parameters have the symmetry  $t_{\bm{r}_i-\bm{r}_j,\uparrow}=t^{*}_{\bm{r}_j-\bm{r}_i,\downarrow}$ such that our model can be thought of as an extended Kane-Mele model~\cite{PhysRevLett.95.146802}. 
The density operator is $n_i=\sum_{\sigma} n_{i\sigma}$, where $ n_{i\sigma}=c^{\dagger}_{i\sigma}c_{i\sigma}$, and the 
 on-site chemical potential term $\mu_in_i$  is 0 without displacement field $E$ (i.e., perpendicular electric field).  In the presence of displacement field we use $\mu_i=\pm D/2$ for sites  $i$ on the A and B sublattice respectively, with $D\approx El {(2 p - 1)}$ where $l$ is the layer separation and $p\sim 0.93$ is the sublattice-layer polarization (Fig.~\ref{fig:Haldanesetup}), meaning that $93\%$ of the density of Wannier orbitals of the A (B) sublattice lies in the top (bottom) layer.  A more accurate description involving modificatons to the  free Hamiltonian can be obtained from DFT, but we do not pursue this here as our focus is on correlations via DMRG.

The Coulomb interaction between Wannier orbitals generate multiple 
four-fermi terms (see supplement for more details).   We include an effective dielectric constant $\epsilon$, whose preciseis not easy to establish as it depends on screening from remote bands and core orbitals of all of the atoms in the environment.  As a reference, an MoTe$_2$ monolayer has  $\epsilon \approx 10$ and the hBN substrate has $\epsilon \approx 5$.   Strictly speaking $\epsilon$ could also depend on distance.  Therefore we treat $\epsilon$ as an adjustable  parameter. With this in mind,  the interaction terms take the form
\begin{align} 
H_{\rm int} &= \frac{1}{2} \sum_{i,j}\left[ \rule{0pt}{12pt} V_{ij} n_i n_j \right. 
+ J_{ij}^{z}s_{z,i}s_{z,j}+J^{+-}_{ij}s^{+}_{i}s^{-}_{j} \label{eq:Hint} \\
  &+ \left. 4\sum_{\sigma} A_{i,j,\sigma} c^{\dagger}_{i\sigma} (1-n_i)  c_{j\sigma}    +4 t^{\mathrm{sp}}_{i,j} c^{\dagger}_{i\uparrow} c^{\dagger}_{i\downarrow} c_{j\downarrow} c_{j\uparrow}\right], \nonumber
\end{align}
where all of these interaction terms scale as $1/\epsilon$.  Here the spin operators are  $s_{z,i}=\frac{1}{2}(c^{\dagger}_{i\uparrow}c_{i\uparrow}-c^{\dagger}_{i\downarrow}c_{i\downarrow})$, $s^{+}_{i}=c^{\dagger}_{i\uparrow}c_{i\downarrow}$, $s^{-}_{i}=c^{\dagger}_{i\downarrow}c_{i\uparrow,}$
Only the $V_{ij}$ term (the Hartree piece) is long ranged. For this we project the dual-gated screened Coulomb interaction which in momentum space is $\tilde{V}(\bm{q}) = \frac{e^2\text{tanh}(|\bm{q}| d)}{2\epsilon_0 \epsilon |\bm{q}|}$.   In this study, we consider $d = 66~\mathrm{\AA}$ and $d = 102~\mathrm{\AA}$, which are experimentally feasible.   The remaining interaction terms stem from orbital overlaps and are hence short-ranged.  We will truncate these to the second-nearest neighbors.  
The structure of the Wannier orbitals modifies the nearest-neighbor and on-site components $V_{ij}$ as well as the $J_{ij}$, as detailed in Ref.~\cite{suppm}, but leaves the long-range tails unaffected. (These modifications, which can be substantial, have a topological origin~\cite{PhysRevB.56.12847,PhysRevB.105.L241102}}. Finally, we observe that the Wannier projection also produces the final two assisted- and pair-hopping terms in Eq.~\eqref{eq:Hint}.

The full Hamiltonian is then given by the sum of the hopping and interactions, 
\begin{equation}
  H = H_0 + H_{\rm int}   
  \label{eq:H}
\end{equation}
If the system is fully spin-polarized due to interactions, then the  extended Kane-Mele model reduces to a Haldane honeycomb model and our Hamiltonian is then 
\begin{align}\label{eq:HFP}
H_{\mathrm{FP}}=& \sum_{i,j} t_{\bm{r}_i-\bm{r}_j,\uparrow} c^{\dagger}_{i\uparrow}c_{j\uparrow}+\sum_{i}\mu_in_{i\uparrow} 
+\frac{1}{2} \sum_{i,j}V^{\rm eff}_{ij} n_{i\uparrow}n_{j\uparrow}.
\end{align}

\begin{figure}[t!]
\begin{center}
\includegraphics[width=0.9\columnwidth]{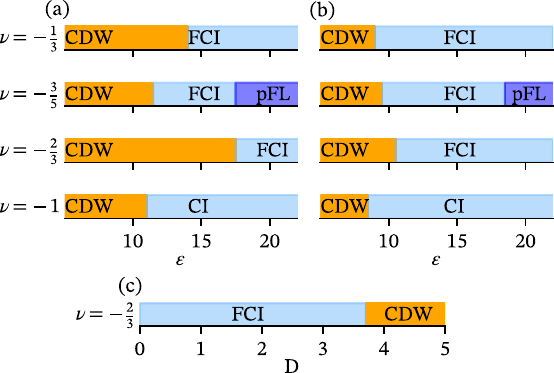}
\end{center}
\caption{
Phase diagrams inferred from YC6/YC5 geometry for screening length (a) $d=66\mathrm{\AA}$ and (b) $d=102\mathrm{\AA}$ as a function of relative dielectric constant $\epsilon$; and (c) sublattice-tuned transition at $\nu=-2/3$ for $d=102\mathrm{\AA}$ and $\epsilon=17$. The pFL stands for spin-polarized  Fermi-liquid. For $|\nu<1|$, all ground states are polarized.}\label{fig:PD}
\end{figure}
\noindent\textit{Numerical Methods.---} 
We primarily use the two-site DMRG algorithm to compute variational ground states of the Hamiltonian Eq.~\eqref{eq:H} on infinite cylinder geometries, but also work with Eq.~\eqref{eq:HFP}  to accelerate calculations when full polarization has been established in the former.
Exact ground states are approached by increasing the number of variational parameters, characterized by the bond dimension $\chi$, while maintaining  a truncation error less than $10^{-5}$.
The maximal width of the cylinders is limited by computational resources.
In most calculations, we choose a  width-6 zigzag nanotube geometry (see Fig.~\ref{fig:Haldanesetup}(c) ``YC6'' in the nomenclature of Ref.~\cite{YanHuseWhite}) for fillings $\nu=-1/3,-2/3, -1$ and YC5 cylinders for  fillings $\nu=-3/5$. Here, the filling $\nu$ is defined as minus the number of holes per unit cell, and the choice of geometry is to allow competing CDWs to be commensurate. We must take care to properly handle the long-range interactions, as follows. First, we employ the standard practice of ensuring that all screening lengths considered are small compared to the cylinder width~\cite{PhysRevB.91.045115}. Second, although we eventually impose a hard cutoff on the interaction at long distances, we keep terms up to as large as 20th and 50th nearest neighbors respectively for the two choices of screening length, to ensure that we capture the essence of the long-range Coulomb tails without artefacts. The uniform-density representation~\cite{suppm} of FCIs on infinite cylinders enables a straightforward distinction between FCI and CDW.  Beyond this, we also identify FCIs via Hall conductance calculations~\cite{suppm}.

\noindent\textit{Phase Diagram.---}
Fig.~\ref{fig:PD} summarizes our DMRG  ground-state phase diagrams of Hamiltonian Eq.~\eqref{eq:H}    at various fillings for (a)  screening length $d=66\mathrm{\AA}$ ($1.28a$) and (b) $d=102\mathrm{\AA}$ ($1.969a$ constant); we also find (c) a  displacement field-tuned FCI-CDW phase transition at $\nu=-2/3$ and $\epsilon=17$, consistent with experiments \cite{PhysRevX.13.031037}, with the CDW a potential candidate for the observed topologically trivial state across the transition. Note that while the $d=102\mathrm{\AA}$ data are the best proxy for the large $d$ case relevant to experiment, the $d=66\mathrm{\AA}$ data is also valuable as  an estimate of the effects of $d$ extrapolation: comparing the enhanced extent FCI regions for the larger $d$ values suggests that increasing $d$ is a fruitful route to strengthening their stability. Comparing results for different $d$ also hints at possible phase transitions that can be accessed by adjusting the degree of  screening, under the assumption that the underlying bands from which $H$ is constructed are relatively insensitive to $d$. Overall, the relative difficulty in stabilizing FCIs for small $d$ suggests that the energetics in tMoTe$_2$ may differ significantly from  those captured by short-ranged interacting  models~\cite{parameswaran2013fractional} for FCIs on the honeycomb lattice~\cite{PhysRevLett.106.236804,PhysRevLett.112.126806,GMZP,PhysRevB.107.L201109}.  We note that strong ferromagnetism and FCIs are absent for $\nu=-1/3$ in experiments, but they nevertheless remain significant features of Fig.~\ref{fig:PD}. We now discuss further details of DMRG data, in particular addressing the ferromagnetism, competing CDWs, and the comparison between $\nu = -1/3$ and $\nu = -2/3$.
 
\begin{figure}[t!]
\begin{center}
\includegraphics[width=0.9\columnwidth]{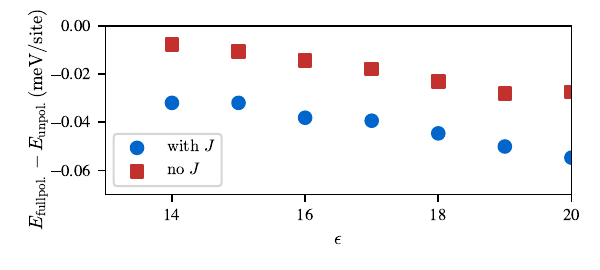}
\end{center}
\caption{The role of exchange interactions in stabilizing ferromagnetism at $\nu=-2/3$ ($\epsilon ~\sim 10$). We show the ground-state energy density difference on YC6 cylinders between the unpolarized (unpol.) sector and fully polarized (fullpol.) sector where we either retain (blue circles) or neglect  (red squares) the exchange interactions in Eq.~\eqref{eq:H}.} 
\label{fig:FPvsNP}
\end{figure}

\noindent\textit{Ferromagnetism.---}
For $1/3<|\nu|\leq 1$, we find that ferromagnetism is stabilized in a range of $\epsilon$ consistent with experiments~\cite{cai_signatures_2023,park2025observationhightemperaturedissipationlessfractional}. However, a key question remains as to the origin of the ferromagnetism, with three salient possibilities:  Stoner-type physics associated with the flatness of the bands, superexchange via pure density-density interactions at finite doping, or direct spin exchange from the local Wannier overlaps, which favours out-of-plane ferromagnetism since $J_{ij}^{z}<0, |J_{ij}^{z}|>J_{ij}^{\pm}>0$ for the smallest distances between $i$ and $j$. To address this question, in Fig.~\ref{fig:FPvsNP} we compare the competition between fully spin-polarized and unpolarized states at $\nu=-2/3$ with and without direct spin exchange interactions, finding a sharp difference between these cases. Since the dispersion and superexchange physics are likely to remain unaffected by this (and we have separately verified that assisted and pair hopping terms minimally influence ferromagnetism for $|\nu|<1$), we conclude that the direct exchange is crucial in lending energetic stability to the ferromagnetic order believed to be present in tMoTe$_2$ for the estimated $\epsilon \sim 10$. (This is indirectly corroborated by our analysis of the importance of band mixing at $\nu=-2/3$, since direct exchange is only possible in the two-band model.)

We also find that FCIs are relatively stronger ferromagnets than their competing CDWs, even though the latter require stronger interactions. For example, for $\epsilon=8$, the ferromagnetic $\nu=-1/3$ CDW only has $\sim 0.007~\text{meV}$ per site lower energy than its unpolarized counterpart. This suggests that ferromagnetic order in the  CDWs is more thermally fragile, which is a potential explanation for the lack of spin polarization coincident with the absence of an FCI at $\nu=-1/3$. This is intuitively consistent with a direct exchange-dominated mechanism for ferromagnetism: since direct exchange is only sizeable for nearest and next-nearest neighbors, it is suppressed for dilute CDWs.

As an aside, we also comment on the magnetization for  $1<|\nu|<2$, a filling range that has drawn much recent attention due to experimental hints of a novel correlated state~\cite{wang2025hidden,xu2025signatures,kwan2024abelianfractionaltopologicalinsulators}. While detailed investigation lies outside the scope of the present work, a preliminary study for $\nu=-4/3$ indicates that the ground state is unpolarized~\cite{suppm}, consistent with experiments~\cite{park2024ferromagnetism,wang2025hidden,xu2025signatures}.

\begin{figure}[t!]
\begin{center}
\includegraphics[width=\columnwidth]{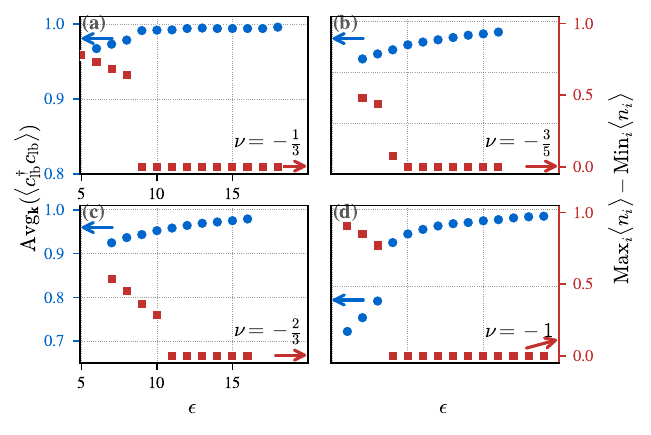}
\end{center}
\caption{Lower-band occupations ($\mathrm{Avg}_{\mathbf{k}}(\langle c^{\dagger}_{\mathrm{lb}}(\mathbf{k}) c_{\mathrm{lb}}(\mathbf{k})) \rangle$) (in blue) and  charge orders  ($\mathrm{Max}_i \langle n_i\rangle-\mathrm{Min}_i \langle n_i\rangle$) (in red) as a function of relative dielectric constant $\epsilon$ for screening length $d=102~\mathrm{\AA}$. Lower-band occupation is one in the case of no band mixing. In order to admit the relevant CDWs, we used the YC6 geometry in (a,c,d) and YC5 in (b).}\label{fig:bandmixingchargeorder}
\end{figure}

\noindent\textit{Competing CDWs and Band Mixing.---} Since the gap between the two lowest bands of tMoTe$_2$ is comparable to the interactions, substantial band mixing is possible. This is allowed in our lattice model, in contrast to single-band-projected Hamiltonians, and is particularly salient near 3.7\degree~ twist angles where adjacent bands have opposite Chern numbers, distinguishing the problem from conventional Landau-level mixing. Indeed, the emergence of charge order as a competitor to FCIs/CIs occurs at small $\epsilon$ where we expect such band mixing to be strong. Accordingly, we now explore the relationship between band mixing and charge order.

To that end, Fig.~\ref{fig:bandmixingchargeorder} shows the occupation of the lower band alongside the site-to-site charge imbalance (a measure of CDW order) at various fillings for $d=102~\mathrm{\AA}$ as a function of $\epsilon$. We find a striking difference between the data at $\nu=-1/3,-1$ when compared to $\nu=-2/3, -3/5$: whereas in the first case the onset of charge order occurs simultaneous with an abrupt increase in band mixing, in the latter cases the two are uncorrelated. Focusing specifically on FCIs, we note that while the $\nu=-2/3, -3/5$ states appear robust against substantial ($\sim 10\%$) band mixing, the $\nu=-1/3$ FCI only survives in regimes where such mixing is negligible.

We now attempt to rationalize some of these features.  First, we note
that $\nu=-1$ is a filled Chern band.  While interaction (particularly
its strong short-range component) causes some band mixing in the
ground state, this just effectively renormalizes the states in the
filled band.  The single-particle gap leaves no freedom to reposition
particles within the filled band and form a charge-ordered state.
Thus, the charge ordering transition that occurs at large interaction
strength can only occur if there is a large increase in band mixing
such that the upper band becomes a near-equal player.

In the FCI phases, some amount of band mixing occurs and renormalizes
the effective interactions within the lower band, but does not change
the fundamental physics of the FCI.  The amount of band mixing is
higher for the higher densities $\nu=-3/5$ and $\nu=-2/3$, simply
because the particles are closer together than for the case of
$\nu=-1/3$, and thereby experience larger band-mixing interactions.

For the CDW phases, however, there is a fundamental difference in how
band mixing affects $\nu=-1/3$ compared to $\nu=-2/3$ (and $\nu=-3/5$
which is similar to $-2/3$).  While naively it might seem like
$\nu=-2/3$ should be similar to $\nu=-1/3$ given that the lower band
is nearly particle-hole symmetric~\cite{PhysRevLett.132.036501,PhysRevLett.127.246403,PhysRevResearch.5.L012015,PhysRevB.109.045147}, in fact this is
not the case.  The CDW at $\nu=-1/3$ is very Wigner crystal-like with
charge lumps well localized fairly far apart from each other (note the
large charge ordering parameter in Fig.~\ref{fig:bandmixingchargeorder}a).  In this case, band
mixing allows the charge lumps to localize more tightly, thereby
further stabilizing the CDW state.  Indeed, it is this band mixing
which favors the $\nu=-1/3$ Wigner-crystal CDW state and causes a
sudden increase in band mixing at the FCI-CDW transition.

In contrast, $\nu=-2/3$ has a more mild charge modulation (Fig.~\ref{fig:bandmixingchargeorder})
rather than the stronger modulation of a point-charge-like~\cite{Kaushal2022} Wigner
crystal.  The particles at filling $-2/3$, since they have double the
density compared to $-1/3$, have to be much closer together and so the
picture of well separated charge lumps no longer applies.  While one
could think of $\nu=-2/3$ as being the particle-hole conjugate of
$\nu=-1/3$, this is true only in the absence of band mixing: one
cannot even describe band mixing in the language of these conjugated
particles.  Thus we are stuck describing the $-2/3$ state as
consisting of charges at higher density that are not well-separated,
which cannot form a point-like Wigner crystal and are therefore
not easily stabilized by increased band mixing.

We also comment on the FCI at $\nu=-2/3$, which is the most prominent fractional filling seen in experiments. Here, while band mixing is apparently less significant to the CDW competition, it remains fairly significant at $\epsilon$ and $d$ consistent with experiment, meaning that the popular single-band lowest Landau level analogy~\cite{PhysRevB.85.241308,PhysRevB.90.165139,PhysRevLett.127.246403,PhysRevB.108.205144,10.21468/SciPostPhys.12.4.118,PhysRevResearch.5.L012015,PhysRevLett.111.126802,PhysRevLett.132.096602,PhysRevB.111.125122,PhysRevLett.134.136502} may be insufficient to explain its stability. Indeed, our phase diagram in Fig.~\ref{fig:PD} shows that the $\nu=-2/3$ FCI shrinks significantly with enhanced screening, suggesting that that the stability of FCIs with realistic band mixing (as captured by our numerics) may depend significantly on the range of interactions, which could potentially be explored by adjusting the screening in experiments.

We relegate further discussion of the various charge ordered states to the Supplement~\cite{suppm}.

\noindent\textit{Nature of the CDW-FCI Transitions.---} As we have already noted, the finite-width infinite cylinder geometry deployed in DMRG is often a very good approximation of the true two-dimensional limit, as quantified by comparing the correlation length to the half-width of the cylinder. Our data~\cite{suppm} shows that the FCI states have a typical correlation length of $\sim 2$ unit cells, so that they satisfy this criterion for the YC5/YC6 cylinder used in our study. However, the correlation length can grow (up to cutoffs imposed by system size or bond dimension) at continuous transitions between the FCI and the competing CDWs. From the charge order data (Fig.~\ref{fig:bandmixingchargeorder}), we conclude that the the $\epsilon$-tuned transitions for $\nu=-2/3,-3/5$ appear to be weakly first-order. However, even in these cases, the CDW correlation length is sufficiently large ($\sim 4$ unit cells) that the DMRG may not be fully reliable. The situation is yet more challenging at the experimentally important $\nu=-2/3$ filling, where the transition is closest to appearing continuous, and we cannot reliably access the exact transition due to finite size effects. We note that experiments~\cite{PhysRevX.13.031037} find a displacement-field-tuned transition from FCI to CDW (cf. Fig.~\ref{fig:PD}(c)), suggesting that this is an important avenue for future work.

\noindent\textit{Discussion.---}     In this work we have examined the energetics of the competing phases in tMoTe$_2$.    In order to allow charge ordered states to arise, and to suppress finite size effects, we have worked with large system sizes accessible only in DMRG.  We examine how the system evolves as a function of screening length, displacement field, and interaction strength (effective dielectric constant $\epsilon$) at a number of different fillings. 

We identified two pieces of physics  that can tip the energy balance of the competing phases.    First, we found that  proper treatment of the  short-ranged spin exchange interaction plays a crucial role in stabilizing spin polarized phases.     Second, we found that proper treatment of band mixing (often neglected for computational expedience) favors charge-ordered phases.   Interestingly, for certain fillings ($\nu=-2/3, -3/5$), we found that the amount of band-mixing appears to be essentially decoupled from the phase transition to charge order as a function of interaction strength ($\epsilon$), suggesting that this transition could be studied by integrating out the higher band first.   However at other fillings, $\nu=-1/3, -1$, there is a discontinuity in the amount of band-mixing at the transition, which potentially makes the transition fundamentally a two-band problem.   

\noindent\textit{Acknowledgements.---}
We thank E.~Bergholtz,  N.~Bultinck, V.~Crépel, X.~Dai, D.~Kennes, J.~ Li, F.~Pollmann, N.~Regnault, R.~Mong, A.~Stern, K.~Vasiliou, C.~Wang, J.~Wang, Z.~Wang, F.~Wu, A.~Seidel, D.~Xiao, K.~Yang and Y.~Zhang for discussions. We are especially grateful to Nicolas Regnault for comments on the manuscript and an illuminating conversation on the interplay of band mixing and charge order. We acknowledge support from the European Research Council (ERC) under the European Union Horizon 2020 Research and Innovation Programme (Grant Agreement Nos. 804213-TMCS; YH, SAP) and from a UKRI Frontier Research Grant EP/Z002419/1(SAP).   Numerical calculations are based on the TeNPy package~\cite{10.21468/SciPostPhysLectNotes.5,iDMRGpaper}, and made use of the  Advanced Research Computing (ARC) facility at the University of Oxford.

\bibliography{bib}

\clearpage
\onecolumngrid
\section{Supplemental Material}
\renewcommand{\theequation}{S\arabic{equation}}
\setcounter{equation}{0}
\renewcommand{\thefigure}{S\arabic{figure}}
\setcounter{figure}{0}
\renewcommand{\thetable}{S\arabic{table}}
\setcounter{table}{0}
\subsection{Parameters of the Hamiltonian}

\begin{figure}[h!]
\begin{center}
\includegraphics[width=0.5\columnwidth]{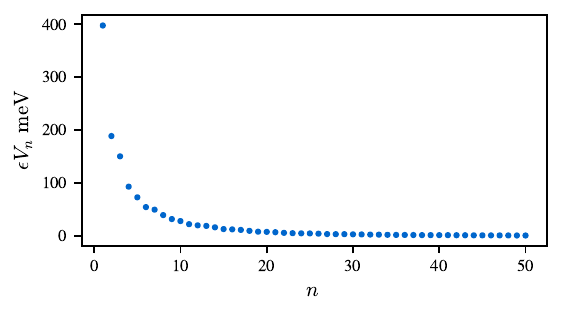}
\end{center}
\caption{Density-density interactions between sites on the lattice for separations ranging from first to  fiftieth neighbour, for a screening length $d=102 \mathrm{\AA}.$ }\label{fig:Vn}
\end{figure}

We summarize the parameters that we use for screening length $d=102\mathrm{\AA}$ in Tab.~\ref{tab:Hs} and Fig.~\ref{fig:Vn}. In the calculations, we have checked  that $t_4$, $t_5$ do not alter the results with the level of precision we have used, and so for the present data taken at the largest bond dimension ($\chi$), we only keep $t_1$, $t_2$ and $t_3$. We have also checked that various other dropped terms (not listed) are also negligible.
\begin{table}[h]
    \centering
    \begin{tabular}{|c|c|c|c|c|c|c|c|c|c|c|c|}
        \hline
        $t_1$ & $t_2$ & $t_3$ & $t_4$  & $t_5$ & $\epsilon V_0$  & $\epsilon J^z_1$  & $\epsilon J^{\pm}_1$ & $\epsilon J^z_2$ & $\epsilon J^{\pm}_2$& $\epsilon A_1$& $\epsilon t^{\mathrm{sp}}_1$\\
        \hline
        -3.855 & -0.977-1.633i & 0.914 & -0.33 &  -0.121 & 1320 & -35 &  8 & -6 & 1+2i & 7
        & 9\\
        \hline\end{tabular}
    \caption{Hamiltonian parameters. Subscripts 0,1...,5 represent the on-site, nearest...,fifth nearest neighbor terms, $t$, $V$, and $J$ denote hopping, charge, and spin-exchange interactions respectively, and all energies are measured in meV. The numbers are shown for $d=102 \mathrm{\AA}$.}
    \label{tab:Hs}
\end{table}

\subsection{Interaction terms from Wannier orbital projection}
In this section, we derive the interaction terms in the Wannier basis.  We start from the continuum model (BM model) with parameters fitted from the DFT data. Note that in the BM approach, the wave functions are taken to be two-dimensional, neglecting the orbital size in the third dimension. When writing down the interactions, we must therefore explicitly account for the fact that the Wannier orbitals are highly ($\sim 93\%$) sublattice-layer polarized.  We set the layer separation as $l=6\mathrm{\AA}$, but the precise value has little influence on the results as long as it remains much smaller than the moir\'e lattice spacing $a_M\approx 51.8~\mathrm{\AA}$. The electron-electron interactions in the continuum theory are standard density-density interactions for a dual-gate screened Coulomb interaction $V_{\mathrm{d}}(r)$. Taking  the bilayer structure into account, we have \begin{equation} 
H_{\rm int} = \int d^2\mathbf{r} d^2\mathbf{r}' \frac{1}{2}V_{s,s'}(\mathbf{r}-\mathbf{r}')  \psi^{\dagger}_{\sigma s}(\mathbf{r})\psi^{\dagger}_{\sigma s'}(\mathbf{r}')\psi_{\sigma s'}(\mathbf{r}')\psi_{\sigma s}(\mathbf{r}),\end{equation}\label{eq:Clint}
where $V_{s,s'}(\mathbf{r}-\mathbf{r}')=V_{\mathrm{d}}(\sqrt{|\mathbf{r}-\mathbf{r}'|^2+l^2(1-\delta_{s,s'})})$ and $\sigma$ and $s$ denote spin and layer indices respectively. We define the Wannier functions $\phi^{s}_{\sigma}(\mathbf{r};\mathbf{R}_i)$  to be localized near the site $i$. Truncating the Hilbert space via $\psi_{\sigma s}(\mathbf{r}) \approx \sum_{i} \phi^{s}_{\sigma}(\mathbf{r};\mathbf{R}_i)c_{i\sigma}$, the projected interaction terms are given by
\begin{align}
H_{\rm int} =\int d^2\mathbf{r} d^2\mathbf{r}' \sum_{i,j,k,l,\sigma,\sigma',s,s'} \frac{1}{2}V_{s,s'}(\mathbf{r}-\mathbf{r}') \phi_{\sigma}^{s*}(\mathbf{r};\mathbf{R}_{i})\phi_{\sigma'}^{s'*}(\mathbf{r}';\mathbf{R}_{j})\phi_{\sigma'}^{s'}(\mathbf{r}';\mathbf{R}_k)\phi_{\sigma}^{s}(\mathbf{r};\mathbf{R}_l)c^{\dagger}_{i\sigma} c^{\dagger}_{j\sigma'} c_{k\sigma'} c_{l\sigma} 
\end{align}
Because of the localization of  the Wannier functions, the only long-range interactions  between orbitals correspond to those where $i=l,j=k$, 
\begin{align}
H_{\mathrm{int,Hartree}}=\int d^2\mathbf{r} d^2\mathbf{r}' \sum_{i,j,\sigma,\sigma',s,s'}\frac{1}{2}V_{s,s'}(\mathbf{r}-\mathbf{r}') \rho^{s}(\mathbf{r};\mathbf{R}_{i})\rho^{s'}(\mathbf{r}';\mathbf{R}_{j})c^{\dagger}_{i\sigma} c^{\dagger}_{j\sigma'} c_{j\sigma'} c_{i\sigma}, 
\end{align}
where $\rho^{s}(\mathbf{r};\mathbf{R}_{i})=\phi_{\uparrow}^{s*}(\mathbf{r};\mathbf{R}_{i})\phi_{\uparrow}^{s}(\mathbf{r};\mathbf{R}_{i})=\phi_{\downarrow}^{s}(\mathbf{r};\mathbf{R}_{i})\phi_{\downarrow}^{s}(\mathbf{r};\mathbf{R}_{i})$ is the charge density of the corresponding Wannier orbital.
Two-site interactions can also be obtained by setting $i=k,j=l,i\neq j$.
We define
\begin{align}\label{w1}
w_{i,j,\sigma,\sigma'}=\int d^2\mathbf{r} d^2\mathbf{r}' \sum_{s,s'}\frac{1}{2}V_{s,s'}(\mathbf{r}-\mathbf{r}')\phi_{\sigma}^{s*}(\mathbf{r};\mathbf{R}_{i})\phi_{\sigma'
}^{s'*}(\mathbf{r}';\mathbf{R}_{j})\phi^{s'}_{\sigma'}(\mathbf{r}';\mathbf{R}_i)\phi^{s}_{\sigma}(\mathbf{r};\mathbf{R}_j).
\end{align}
We then have $w_{i,j,\sigma,\sigma'}^{*}=w_{i,j,\sigma',\sigma}$, $w_{j,i,\sigma,\sigma'}^{*}=w_{i,j,\sigma,\sigma'}$ and it follows that $w_{i,j,\uparrow,\uparrow}=w_{i,j,\downarrow,\downarrow}$, allowing us to write the first exchange contribution as
\begin{align}\label{exchange}
H_{\mathrm{int, exchange 1}}=-\sum_{i,j,\sigma}w_{i,j,\sigma,\sigma} c^{\dagger}_{i\sigma} c_{i\sigma} c^{\dagger}_{j\sigma} c_{j\sigma} -\sum_{i,j} w_{i,j,\uparrow,\downarrow} c^{\dagger}_{i\uparrow} c_{i\downarrow} c^{\dagger}_{j\downarrow} c_{j\uparrow}-\sum_{i,j} w_{i,j,\downarrow,\uparrow} c^{\dagger}_{i\downarrow} c_{i\uparrow} c^{\dagger}_{j\uparrow} c_{j\downarrow}. 
\end{align}
Next, using $\sum_\sigma c^{\dagger}_{i,\sigma} c_{i,\sigma} c^{\dagger}_{j,\sigma} c_{j,\sigma}=n_in_j/2+2s_{z,i}s_{z,j}$, $s^{+}_{i}=c^{\dagger}_{i\uparrow}c_{i\downarrow}$, $s^{-}_{i}=c^{\dagger}_{i\downarrow}c_{i\uparrow}$, we can rewrite Eq.~\eqref{exchange} as 
\begin{align}
H_{\mathrm{int, exchange 1}}=\sum_{ij}\left[\frac{1}{2}J_{ij}^{z}s_{z,i}s_{z,j}+\frac{1}{2}J^{+-}_{ij}s^{+}_{i}s^{-}_{j}-\frac{1}{2}w_{i,j,\uparrow,\uparrow} n_in_j\right],
\end{align}
where $J_{ij}^z=-4w_{i,j,\uparrow,\uparrow},J_{ij}^{+-}=-4w_{i,j,\uparrow,\downarrow}$.
The terms above are included in our lattice model in Eq.~\eqref{eq:H}:
\begin{align}\label{eq:Hintredux}
H_{\mathrm{int,Hartree}}+H_{\mathrm{int, exchange 1}}=\sum_{i,j}\left[\frac{1}{2}V(|\bm{r}_i-\bm{r}_j|)n_in_j+\frac{1}{2}J_{ij}^{z}s_{z,i}s_{z,j}+\frac{1}{4}J^{+-}_{ij}s^{+}_{i}s^{-}_{j}\right],
\end{align}
where $V(|\bm{r}_i-\bm{r}_j|)=\int d^2\mathbf{r} d^2\mathbf{r}' \sum_{s,s'} V_{s,s'}(\mathbf{r}-\mathbf{r}') \rho^{s}(\mathbf{r};\mathbf{R}_{i})\rho^{s'}(\mathbf{r}';\mathbf{R}_{j})-w_{i,j,\uparrow,\uparrow}$. When projecting to the fully polarized sector, $H_{\mathrm{int,Hartree}}+H_{\mathrm{int, exchange 1}} \rightarrow \sum_{i,j}\frac{1}{2}V^{\mathrm{eff}}(|\bm{r}_i-\bm{r}_j|)n_{\uparrow,i}n_{\uparrow,j}$, where $V^{\mathrm{eff}}=V(|\bm{r}_i-\bm{r}_j|)-w_{i,j,\uparrow,\uparrow}$.For small $|\nu|$,other two-site terms are less important as the on-site repulsion largely suppresses the double occupancy.

The two-site $s$-wave pair hopping terms can be obtained by setting $i=j,k=l$. We define
\begin{align}
t^{\mathrm{sp}}_{i,j}=\int d^2\mathbf{r} d^2\mathbf{r}' \sum_{s,s'}\frac{1}{2}V_{s,s'}(\mathbf{r}-\mathbf{r}')\phi_{\uparrow}^{s*}(\mathbf{r};\mathbf{R}_{i})\phi_{\downarrow
}^{s'*}(\mathbf{r}';\mathbf{R}_{i})\phi^{s'}_{\downarrow}(\mathbf{r}';\mathbf{R}_j)\phi^{s}_{\uparrow}(\mathbf{r};\mathbf{R}_j)
\end{align}
which we can relate to the previously computed $w$ function via $t^{\mathrm{sp}}_{i,j}=w_{i,j,\uparrow,\uparrow}$. The pair hopping term is then given by
\begin{align}
H_{\mathrm{int, exchange 2}}=\sum_{i,j} 2 t^{\mathrm{sp}}_{i,j} c^{\dagger}_{i\uparrow} c^{\dagger}_{i\downarrow} c_{j\downarrow} c_{j\uparrow}.
\end{align}
The two-site assisted hopping terms can be obtained by setting three of $i,j,k,l$ to be equal; we define
\begin{align}
A_{i,j,\sigma}=\int d^2\mathbf{r} d^2\mathbf{r}' \sum_{s,s',\sigma'}\frac{1}{2}V_{s,s'}(\mathbf{r}-\mathbf{r}')\phi_{\sigma}^{s*}(\mathbf{r};\mathbf{R}_{i})\phi_{\sigma'
}^{s'*}(\mathbf{r}';\mathbf{R}_{i})\phi^{s'}_{\sigma'}(\mathbf{r}';\mathbf{R}_i)\phi^{s}_{\sigma}(\mathbf{r};\mathbf{R}_j),
\end{align}
in terms of which we have 
\begin{align}\label{ah}
H_{\mathrm{int, exchange 3}}=\sum_{i,j,\sigma} 2A_{i,j,\sigma} c^{\dagger}_{i\sigma} n_i  c_{j\sigma} 
\end{align}
However, Eq.~\eqref{ah} is  incorrect because the DFT band structure is derived for fully filled valence bands. That justifies working with holes from the outset, namely,  $H_{\rm int} = \int d^2\mathbf{r} d^2\mathbf{r}' \frac{1}{2}V_{s,s'}(\mathbf{r}-\mathbf{r}')  \psi_{\sigma s}(\mathbf{r})\psi_{\sigma s'}(\mathbf{r}')\psi^{\dagger}_{\sigma s'}(\mathbf{r}')\psi^{\dagger}_{\sigma s}(\mathbf{r})$ instead of Eq.~\eqref{eq:Clint}. This does not affect any other terms, but Eq.~\eqref{ah} is modified as \begin{align}\label{ahm}
H_{\mathrm{int, exchange 3}}=\sum_{i,j,\sigma} 2A_{i,j,\sigma} c^{\dagger}_{i\sigma} (1-n_i)  c_{j\sigma} 
\end{align}
In other words, this term is expected to not significantly modify the dispersion near the case of fully filled valence bands ($c_{i,\sigma}^{\dagger}c_{i,\sigma} \approx 1$).
Indeed, as long as one species of spin is fully filled, i.e.,$c_{i,\downarrow}^{\dagger}c_{i,\downarrow}=1$ , we have  
\begin{align}
c_{i,\uparrow}^{\dagger}(1-n_i)c_{j,\uparrow}= c_{i,\uparrow}^{\dagger}(1-c_{i,\uparrow}^{\dagger}c_{i,\uparrow}-c_{i,\downarrow}^{\dagger}c_{i,\downarrow})c_{j,\uparrow}=0
\end{align}
Terms that involve three or four sites have small enough amplitudes and thus are neglected in the modeling. 

\subsection{Band projections}

In the main text, we present data on lower-band occupations to quantify the band mixing. This quantity can be extracted from lattice correlation function, by relating the lower-band occupation operator to lattice operators:
\begin{align}
\frac{1}{N}\sum_{k_x,k_y} c_{k_x,k_y,1}^{\dagger}c_{k_x,k_y,1} &= \frac{1}{N^2}\sum_{k_x,k_y,x,x',y,y',s,s'} e^{ik_xx+ik_yy} U^{*}_{1,s}(k_x,k_y) c_{x,y,s}^{\dagger} e^{-ik_xx'-ik_yy'}U_{1,s'}(k_x,k_y)(k_x,k_y)c_{x',y',s'}  \nonumber \\
&= \frac{1}{N}\sum_{x,y,x',y',s,s'} R_{s,s'}(x-x',y-y') c_{x,y,s}^{\dagger} c_{x',y',s'}, 
\end{align}
with
\begin{align}\label{Rdef}
R_{s,s'}(\Delta x,\Delta y)
= \frac{1}{N} \sum_{k_x,k_y} e^{i(k_x\Delta x+k_y \Delta y)} U^{*}_{1,s}(k_x,k_y) U_{1,s'}(k_x,k_y),
\end{align}
where $e^{(ik_xx+ik_yy)}U_{1,s}(k_x,k_y)$ is the Bloch function of the lower band (labeled as $1$) as a function of sublattice index $s$ and $N$ is the total number of sites. Due to translational invariance, the division by $N$ is equivalent to the average over a unit cell. It turns out that $R_{s,s'}(\Delta x,\Delta y)$ is given by the ground-state correlation function of the model without interactions:
For a (noninteracting) band insulator, we have
\begin{align}\label{BIcdc}
&\langle c_{x,y,s}^{\dagger} c_{x',y',s'} \rangle_{\mathrm{BI}} \nonumber\\
=&  \sum_{k_x,k_y} e^{-i(k_x\Delta x+k_y \Delta y)} \langle c_{k_x,k_y,s}^{\dagger} c_{k_x,k_y,s'} \rangle/N \nonumber \\
=&  \sum_{k_x,k_y,b,b'} e^{-i(k_x\Delta x+k_y \Delta y)} \tilde{U}^{*}_{s,b}(k_x,k_y)\tilde{U}_{s',b'}(k_x,k_y) \langle   c_{k_x,k_y,b}^{\dagger} c_{k_x,k_y,b'} \rangle/N \nonumber \\
=&  \sum_{k_x,k_y} e^{-i(k_x\Delta x+k_y \Delta y)} \tilde{U}^{*}_{s,1}(k_x,k_y)\tilde{U}_{s',1}(k_x,k_y)/N,  
\end{align}
where  $\tilde{U}_{s,1}(k_x,k_y)=U^{*}_{1,s}(k_x,k_y)$.

Comparing Eq.~\eqref{Rdef} with Eq.~\eqref{BIcdc}, we find that $ R_{s,s'}(x-x',y-y') =\langle c_{x,y,s}^{\dagger} c_{x',y',s'} \rangle_{\mathrm{BI}}^{*}$

\subsection{More DMRG details}  

\begin{figure}[t!]
\begin{center}
\includegraphics[width=0.5 \columnwidth]{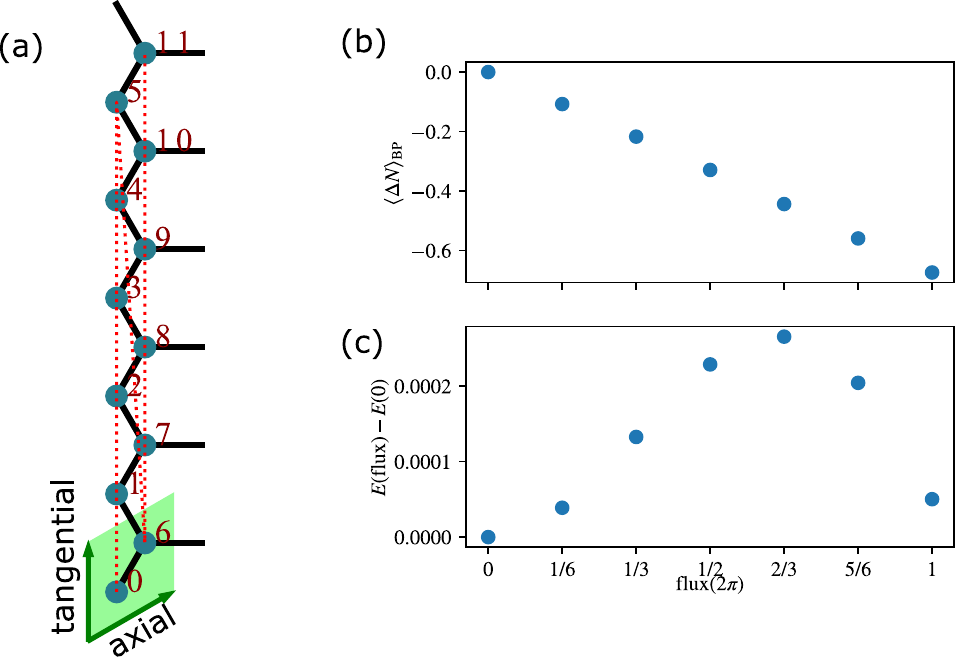}
\end{center}
\caption{YC6 geometry and FCI representative states phase at $\nu=-2/3$  (a) The MPS unit cell and the order of sites (b) Charge pumping via flux insertion (c) Energy per-site change induced by flux insertion. One cycle of flux insertion results in states with different charge polarization which may be not strictly degenerate ground states}\label{fig:YC6nu23}
\end{figure}

\begin{figure}[t!]
\begin{center}
\includegraphics[width=0.5\columnwidth]{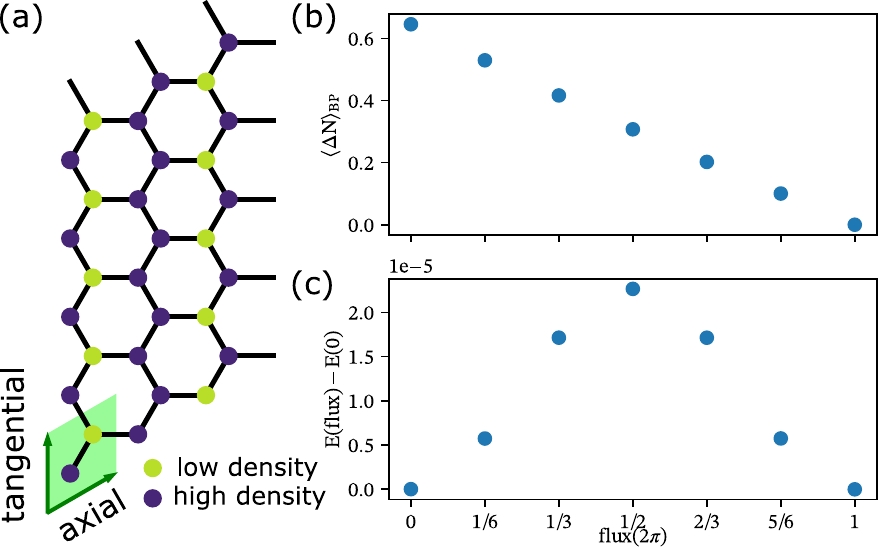}
\end{center}
\caption{YC5 geometry and FCI representative states phase at $\nu=-2/3$  (a) Finite-width-induced artificial CDWs in a unit cell of MPS in DMRG at zero flux. The low and high density difference is smaller than 0.02 for parameter $\epsilon \approx 21, D= 65~\mathrm{AA}$ (b) Charge pumping via flux insertion (c) Energy per-site change induced by flux insertion. One cycle of flux insertion results in degenerate CDWs moved by one lattice constant along the axial direction from (a).}\label{fig:YC5nu23}
\end{figure}

The DMRG algorithm is performed to optimize infinite matrix product states (MPS). By increasing the bond dimension, the optimized infinite MPS can approach the ground states of Hamiltonians in infinite-cylinder geometries. We work with a real space basis. The MPS is defined as follows; sites are indexed in a line, labeled as $i=1,2,...,N$. The order we choose is labeled in Fig.~\ref{fig:YC6nu23} . Each site has a matrix $A_{s_i}$, depending on the local Hilbert-space configuration $s_u$. Each $A_{s_i}$ is a $\chi$ by $\chi$ matrix, where  $\chi$ is the bond dimension. The data shown in the main text are obtained using $\chi=4800$ and all data points are converged with an error bar smaller than the marker size. The MPS is obtained by consecutive contractions of the form $...\left(\prod_{i-1}^N A_{s_i}\right)\left(\prod_{i-1}^N A_{s_i}\right)...$, where the site matrix $i$ is multiplied consecutively and $...$ denotes the repetition of unit cells and is terminated by a fixed-point tensor at infinity.  The set of $i$ determines the MPS unit cell and must include sites on a rung or multiple rungs. The  multiple-site unit cell enforces translational symmetry, as it corresponds to multiples of the lattice constant along the axial direction. The size of this unit cell can be adjusted to search for ground states with different possible translational symmetry breaking patterns along the axial direction. For example, the MPS denoted in Fig.~\ref{fig:YC5nu23}, has a three-site unit cell. 

\begin{figure}[t!]
\begin{center}
\includegraphics[width=0.5\columnwidth]{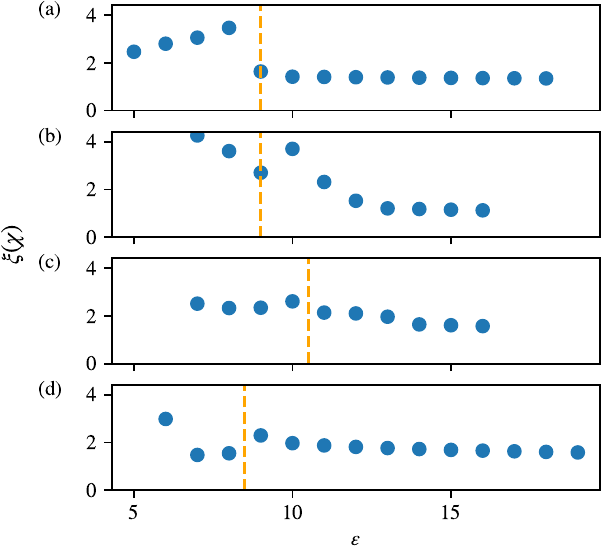}
\end{center}
\caption{DMRG correlation length ($\chi=4800$) as a function of relative dielectric constant $\epsilon$ for screening length $d=102~\mathrm{\AA}$. (a) $\nu=-1/3$ (b) $\nu=-3/5$ (c)  $\nu=-2/3$ (d) $\nu=-1$. Dashed orange line indicates boundary between CDW and FCI/CI phases. Note that the DMRG correlation length near the transition of $\nu=-2/3$ is not converged in $\chi$ and the short correlation length is an artifact.}\label{fig:correlation_length}
\end{figure}

We compute the correlation functions using the (converged) MPS transfer matrix  $T=\sum_{s_1,s_2,...,s_N}\prod_i A^{*}_{s_i} \otimes \prod_i A_{s_i}$.  $T$ has four indices inherited from left indices of $A_1$, $A^{*}_1$ and the right indices of $A_N$, $A^{*}_N$.  Grouping the indices as $T_{a^*,a;b^*,b}$, $T$ can be considered as a matrix. Denote the eigenvalues of $T$ by $\lambda_j$.  Due to normalization and the requirement of injectivity, we have $ \lambda_j \leq 1 $, with the dominant eigenvalue $ \lambda_1 = 1$ being nondegenerate.
 We denote the second largest $|\lambda_i|$ as $|\lambda_2|$. The correlation length of a gapped state is the maximal exponential decay length scale of any two-point correlation function. As the sites are in a line in MPS, the best proxy of the correlation length is $\log (1/|\lambda_2|)/N_{\rm{min}}$, where $N_{\rm{min}}$ is the width of the cylinder. We have to increase $\chi$ to get a better approximation of the ground state.  The estimate $\xi(\chi)$ typically increases with $\chi$. Thus, for a fixed $\chi$, $\xi(\chi)$ generally provides a lower bound.
 Since we implement charge conservation in MPS, $A$s are block diagonalized, as is $T$. Thus we can find the correlation in different charge sectors:" specifically, the diffrent $U(1)$ charges carried by the operators in the two-point correlations. The data from the neutral sector are plotted in Fig.~\ref{fig:correlation_length}. 

Now we discuss representative FCI states in infinite-cylinder geometry.  This has been briefly mentioned in Ref.~\onlinecite{GMZP}. Here we give a detailed discussion based on our data.

The Lieb-Schulz-Mattis (LSM) theorem enforces that any rung must have an integer number of electrons on average to be able to have a ground state with no translational symmetry breaking. This immediately tells us that the representative states for FCIs in the YC5 geometry must break the translational symmetry breaking. Formally , in a fully 2D problem an FCI evades this requirement of symmetry breaking because of its topological ground state degeneracy, but this is precluded in finite cylinders; thus, we need to understand how to diagnose the topological from the evolution of symmetry breaking with cylinder width. The minimal possible symmetry breaking pattern is plotted in Fig.~\ref{fig:YC5nu23} and is what we obtained in DMRG.  The CDW order of the state is very weak ($\sim0.02$) and is a double stripe -CDW. 

That such a state is a representative of an FCI can be diagnosed by its {\it fractional charge pumping} property:  namely, adiabatically inserting a unit of $U(1)$ flux results in a degenerate CDW shifted by one lattice constant along the axial direction. This is to be contrasted to the  ``classical“ CDWs we find at strong coupling, which have trivial charge pumping.  Furthermore, assuming there is no intertwining of the CDW with the `true' topological order, we expect the CDW order to decrease to zero when increasing the cylinder width.  So we view such a CDW with charge pumping  as an artificial order linked to the finite cylinder width. We note that representations of conventional Landau level quantum Hall states on cylinders are also well known to also have artificial CDW order, decaying with cylinder width. There is one difference; however, artificial CDWs for FCI representative states have fixed unit cell size, whereas in the FQHE the unit cell size shrinks.

Finally, we turn to the YC6 geometry representation of $\nu=-2/3$ FCI discussed in the main text. According to LSM, the ground state could be an insulating state without transnational symmetry breaking, which normally cannot have exact ground state degeneracy corresponding to that of FCI in the 2D limit. On the other hand, akin to YC5, representing states with artificial CDW realizing exact degeneracy is also possible.

In DMRG, we get very small CDW order parameters, which could potentially be vanishing if DMRG has perfect accuracy. We conjecture that the representative state of the $\nu=-2/3$ FCI is an insulating state with no symmetry breaking. This raises the question of how the ground state degeneracy of this FCI is realized. Our data indicate that the degeneracy is not exact in the YC6 geometry.  In the DMRG charge pumping,  the state apparently should be the ground state of the Hamiltonian with additional flux. The Hamiltonians with flux 0 and $2\pi$ are the same; however, the charge pumping signature of the FCI representative states requires different charge polarization and therefore should be different ground state. In fact, we find that the state inFig.~\ref{fig:YC6nu23}  with flux $2\pi$ has a slightly higher energy. In contrast, those of $\nu=-2/3$ states on YC5 seem to be exactly degenerate. So we obtain metastable states in DMRG for $2\pi$ in YC6 charge pumping.   Note that the entire process is somewhat challenging to access numerically because it requires DMRG to be stuck in local minimum. This is only possible when the cylinder geometry captures the 2D limit pretty well, such that the splitting between the topologically degenerate states is sufficiently small. 

\begin{figure}[t!]
\begin{center}
\includegraphics[width=0.5\columnwidth]{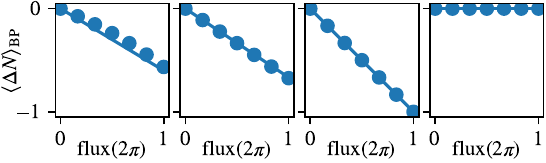}
\end{center}
\caption{
Charge pumping via adiabatic flux insertion. The charge polarization $\langle \Delta N \rangle_{\mathrm{BP}}$ as a function of flux reveals the Hall conductance as its slope. The parameters for each plots are  $\nu=-\frac{3}{5}$, $\nu=-\frac{2}{3}$, $\nu=-1$ with $D=0$, and $\nu=-\frac{2}{3}$ with $D=5~\mathrm{meV}$. The solid lines are a guide for eyes with a slope $-\frac{3}{5}$,$-\frac{2}{3}$,-1, and 0.
}\label{fig:chargepumping}
\end{figure}

\subsection{Additional data for magnetism}
\begin{figure}[t!]
\begin{center}
\includegraphics[width=0.5\columnwidth]{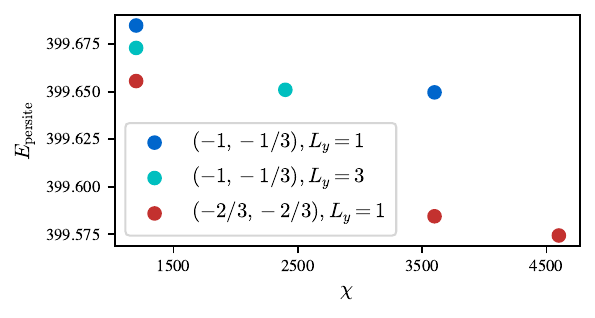}
\end{center}
\caption{Tentative investigation of the magnetic polarization for $\nu=-4/3$. The DMRG energies with bond dimension $\chi$ for partially polarized state (-1,-1/3) and unpolarized state (-2/3,-2/3) are plotted. The screening length is $d=102\mathrm{\AA}$ and dielectric constant is $\epsilon=15$.}\label{fig:polarization43}
\end{figure}

Here we provide preliminary  data for a  study of ferromagneitsm at $\nu=-4/3$ . We consider three different kinds of polarized states. (1) unpolarized $\nu_{\uparrow}=-2/3,\nu_{\downarrow}=-2/3$ (2) partially polarized $\nu_{\uparrow}=-1,\nu_{\downarrow}=-1/3$; and(3) fully polarized $\nu_{\uparrow}=-4/3,\nu_{\downarrow}=0$. Our calculations indicate that the fully polarized case has significantly higher energy relative to the other two cases. This outcome is expected, as heavy doping of the second band leads to an energy cost, determined by the $8$-meV band gap, that exceeds the observed magnetic energy scales.
In Fig.~\ref{fig:polarization43}, we compare the DMRG estimate of the ground-state energy for the unpolarized and partially polarized sectors. With the implementable bond dimension, the truncation error is of the order of $10^{-4}$, and the accuracy is insufficient to definitively fix the quantum phases. Nonetheless, the estimated energy precision is sufficient to infer that the unpolarized state is favored.

\subsection{Charge density waves}
\begin{figure}[t!]
\begin{center}
\includegraphics[width=0.5\columnwidth]{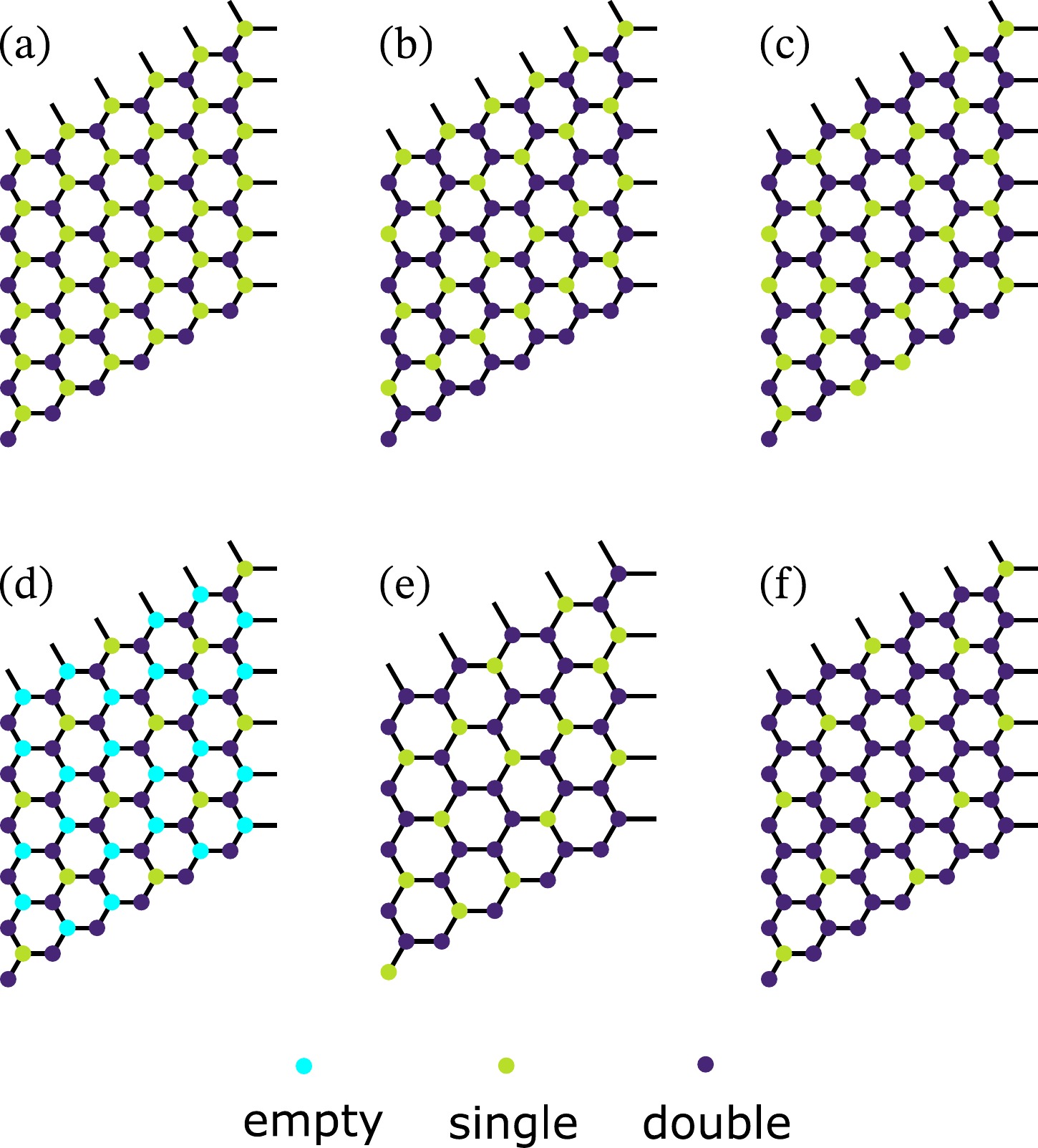}
\end{center}
\caption{Configuration representations of charge density waves obtained in DMRG for various fillings. (a) $\nu=-1$,  (b)(c)(d) $\nu=-2/3$, (e) $\nu=-3/5$, (f) $\nu=-1/3$. In case (d) along, we applied an additional large sublattice potential. The configurations in (b) and (c) are in close energetic competition.}\label{fig:cdws}
\end{figure}

In Fig.~\ref{fig:cdws}, we present the charge-density-wave configurations discussed in the main text. These are depicted as classical configurations to illustrate their symmetries.   For $\nu=-1/3$ and $\nu=-1$, the $C_3$ symmetric classical configurations are believed to minimize potential energies. The energy landscape is less clear for $\nu=-2/3$ and $\nu=-3/5$. In particular, for  $\nu=-2/3$,  low-energy configurations such as Fig.~\ref{fig:cdws} (b)(c) ~\cite{Kaushal2022} are in close competition. When a large sublattice potential ($D$) is introduced in the Hamiltonian, a $C_3$ symmetric configuration of $\nu=-2/3$ —shown in 
Fig.~\ref{fig:cdws} (d)- is believed to minimize the potential energy.
\end{document}